\documentclass[aps,prl,amssymb]{revtex4}

\usepackage{graphicx}
\usepackage{bm}

\def\ep {\epsilon}

\def\e2 {\epsilon-\epsilon_k}
\def\be {\begin{equation}}
\def\ee {\end{equation}}
\def\bea {\begin{eqnarray}}
\def\eea {\end{eqnarray}}

\def\cd {c^{\dagger}}

\def\ua {\uparrow}
\def\da {\downarrow}
\def\si {\sigma}

\def\la {\lambda}
\def\g {\gamma}
\def\ze {\zeta}

\def\de {\delta}

\begin{document}
\title{A fermionic superfluid state for many spinful species - III}

\author{ George Kastrinakis$^*$}

\affiliation{ Institute of Electronic Structure and Laser (IESL), 
Foundation for Research and Technology - Hellas (FORTH),
P.O. Box 1527, Iraklio, Crete 71110, Greece}

\date{October 29, 2012}

\begin{abstract}

In two previous reports, 
we introduced a new fermionic variational wavefunction, 
suitable for interacting multi-species systems and sustaining
superfluidity. This disentangled
wavefunction contains a new quantum index.
Here we introduce a general spin triplet version of this wavefunction.

\end{abstract}

\maketitle

In two previous reports \cite{gk,gk2}, we introduced a new 
variational wavefunction, 
suitable for interacting multi-species systems and sustaining
superfluidity.  This a generalization of the 
Bardeen-Cooper-Schrieffer (BCS) wavefunction \cite{bcs} 
$|\Psi_{\text {BCS}}\rangle  = \prod_{k} 
(u_{k} + v_{k} \; \cd_{k,\ua} \cd_{-k,\da})|0\rangle $. 
The creation/annihilation operators $\cd_{k,\si}/c_{k,\si}$ describe
fermions with momentum $k$ and spin $\si$, and $|0\rangle $ is the vacuum
state. 

We introduce a spin triplet version of this wavefunction, which 
is a generalization of the Balian-Werthamer state \cite{bw}, including
all three components of the total spin. In \cite{gk2} we
introduced the "equal spin pairing" (ESP) case with parallel pair spins
only. 

Let the usual fermionic operators be $\cd_x/c_x$ with $x=\{i,k,\si\}$, 
where $i$ denotes the fermion species/flavor.

We introduce a new quantum index, which is related to
the internal symmetry of the quantum state. It serves to enumerate 
the otherwise "entangled" components of the quantum state
as a function of both momentum and spin. In this way, the treatment
of the coherence factors (coefficients entering the wavefunction
and then everywhere else in the theory) is greatly facilitated.
Thereby we introduce the {\em new fermionic operators}
$\cd_{x,\mu}/c_{x,\mu}$ obeying the anticommutators ($\{a,b\}=ab+ba$)

\be
\{c_{x,\mu},c_{y,\nu}\}=0 \;\;,\;\;
\{c_{x,\mu},\cd_{y,\nu}\}=\de_{xy} \; \de_{\mu\nu}  \;\;.
\ee

Then, we write the usual  $\cd_x/c_x$ as the superposition

\be
\cd_x = \sum_{\de=1}^{N_o} \g_{x,\de}^* \; \cd_{x,\de} \;\; , \;\;
c_x = \sum_{\de=1}^{N_o} \g_{x,\de} \; c_{x,\de} \;\; .  \label{anac}
\ee

The usual anticommutation relations of $\cd_x/c_x$ are preserved,
by imposing the normalization condition 

\be
 \sum_{\de=1}^{N_o} |\g_{x,\de}|^2 = 1 \;\;, \;\; 
\ee

for the weight coefficients $\g_{x,\delta}$, while

\be
\{c_{x},c_{y,\nu}\}=0 \;\;,\;\;
\{c_{x},\cd_{y,\nu}\}=\de_{xy}  \; \g_{x,\nu}  \;\;.
\ee

We also introduce
\bea
C_{i,k,\de}^{\dagger} = u_{i,k} 
+ v_{i,k} \;  \big( \cd_{i,k,\ua,\de} \; \cd_{i,-k,\da,\de}
 + \cd_{i,k,\da,\de} \; \cd_{i,-k,\ua,\de}  \big)
+ w_{i,k,\ua} \; \cd_{i,k,\ua,\de} \; \cd_{i,-k,\ua,\de}
+ w_{i,k,\da} \; \cd_{i,k,\da,\de} \; \cd_{i,-k,\da,\de}  \\
+  s_{i,k} \; \big( \cd_{i,k,\ua,\de} \; \cd_{j,-k,\da,\de}
+ \cd_{i,k,\da,\de} \; \cd_{j,-k,\ua,\de} 
+ \cd_{i,-k,\ua,\de} \; \cd_{j,k,\da,\de}
+ \cd_{i,-k,\da,\de} \; \cd_{j,k,\ua,\de} \big)    \nonumber \\
+ t_{i,k,\ua} \;  \big( \cd_{i,k,\ua,\de} \; \cd_{j,-k,\ua,\de}
+ \cd_{i,-k,\ua,\de} \; \cd_{j,k,\ua,\de}  \big)
+ t_{i,k,\da} \;  \big( \cd_{i,k,\da,\de} \; \cd_{j,-k,\da,\de}
+ \cd_{i,-k,\da,\de} \; \cd_{j,k,\da,\de}  \big)
\;\;.   \nonumber
\eea
$C_{i,k,\de}^{\dagger}$ is a bosonic operator, creating spin triplet
pairs of fermions (for singlet pairs c.f. \cite{gk}),
and $(i,j)=\{(1,2),(2,1)\}$. Other variants of $C_{i,k,\de}^{\dagger}$
can be envisaged as well.

Henceforth we divide the momentum space into two parts, say $k>0$ (sgn$(k)=+$)
and $k<0$ (sgn$(k)=-$).
For $k>0$ we form the following multiplet of $C_{i,k,\de}^{\dagger}$'s
\be
M_k^{\dagger}= C_{1,k,\de=1}^{\dagger} \; C_{2,k,\de=2}^{\dagger} \;\;.
\ee
This multiplet creates all states with momenta $\pm k$, and we take
$N_o=2$ (c.f. \cite{gk,gk2} also).

The new index allows for the bookkeeping
of a superposition of states of a given particle,
i.e. same $x=\{i,k,\si \}$, without the difficulties due to entanglement
within the multiplet $M_k^{\dagger}$, if the index were removed.
In that case, the treatment of the coherence factors $u,v,w,s,t$ 
is prohibitively complicated.

We note that there is {\bf no change} whatsoever implied in the Hamiltonian or 
in the representation of any observable. 

Now we introduce the disentangled state
\be
|\Psi\rangle  = \prod_{k>0} M_k^{\dagger} \; |0\rangle  \;\;.
\ee
Note that {\em all} $C_{i,k,\de}^{\dagger}$'s in $|\Psi\rangle$ {\em commute 
with each other.}

$|\Psi\rangle$ generalizes $|\Psi_{\text {BCS}}\rangle$ and
sustains superfluidity.
This wavefunction makes sense for two or more fermion 
species, with an interaction between different species. It can obviously
be generalized for three or more fermion species.
Moreover, a similar wavefunction using the new quantum index can be written 
in the real space representation instead of the momentum space one.
$|\Psi\rangle$ does represent a very promising avenue for the treatment
of multispecies fermionic systems, as can be seen
from the discussion which follows.

It allows for inequivalence between spin up and down fermions.  
Plus, it allows for the "exact" variational treatment
of a  wider class of Hamiltonians than sheer 
BCS type, e.g. comprising interaction and hybridization 
between different fermion species, in the well 
known manner of the BCS-Gorkov theory \cite{bcs},\cite{agd}.

The normalization $\langle \Psi|\Psi\rangle =1$ implies
\be
|u_{i,k}|^2+2|v_{i,k}|^2+|w_{i,k,\ua}|^2+|w_{i,k,\da}|^2+4|s_{i,k}|^2+
2|t_{i,k,\ua}|^2+2|t_{i,k,\da}|^2=1 \; . \;
\ee
Fermion statistics yields $v_{i,-k}=-v_{i,k}$ and $w_{i,-k,\si}=-w_{i,k,\si}$.

\vspace{.3cm}

{\bf Two fermion species at zero temperature.}
Calculations are straightforward for the matrix elements derived
from $|\Psi\rangle$. E.g. for 2 fermion species with dispersions
$\ep_{i,k,\si}=\ep_{i,-k,\si}$ and for $k>0$ we have
\bea
c_{1,k,\ua} \; M_k^{\dagger} \; |0\rangle = \g_{1 k \ua,1}
(v_{1,k} \; \cd_{1,-k,\da,\de=1} + w_{1,k,\ua} \; \cd_{1,-k,\ua,\de=1} 
+  s_{1,k} \; \cd_{2,-k,\da,\de=1} +  t_{1,k,\ua} \; \cd_{2,-k,\ua,\de=1})
\;  C_{2,k,\de=2}^{\dagger} 
\; |0\rangle \nonumber \\
- \g_{1 k \ua,2} \; (s_{2,k} \;  \cd_{2,-k,\da,\de=2} 
+t_{2,k,\ua} \;  \cd_{2,-k,\da,\de=2} ) \; C_{1,k,\de=1}^{\dagger} \;  
|0\rangle \;\;.
\eea
Then
\be
\langle 0| \; M_k \; \cd_{1,k,\ua} c_{1,k,\ua} \; M_k^{\dagger} \; |0\rangle =
|\g_{1 k \ua,1}|^2 ( |v_{1,k}|^2 + |w_{1,k,\ua}|^2 + |s_{1,k}|^2
+|t_{1,k,\ua}|^2)
+ |\g_{1 k \ua,2}|^2 (|s_{2,k}|^2  +|t_{2,k,\ua}|^2) \;\;. \;\;
\ee
Likewise,
\be
\langle 0| \; M_k \; \cd_{2,k,\ua} c_{1,k,\ua} \; M_k^{\dagger} \; |0\rangle =
-\g_{1 k \ua,1}\; \g_{2 k \ua,1}^* \;
(t_{1,k,\ua}^* \;w_{1,k,\ua} + v_{1,k}\; s_{1,k}^*) 
-\g_{1 k \ua,2}\; \g_{2 k \ua,2}^* \;
(t_{2,k,\ua}\;w_{2,k,\ua}^* + v_{2,k}^* \;s_{2,k} )  \;\;.\;\;
\ee
and
\be
\langle 0| \; M_k \; c_{2,-k,\ua} c_{1,k,\ua} \; M_k^{\dagger} \; |0\rangle =
\g_{1 k \ua,1}\; \g_{2 -k \ua,1} \; u_{1,k}^* \;  t_{1,k,\ua}
-  \g_{1 k \ua,2}\; \g_{2 -k \ua,2} \; u_{2,k}^* \;  t_{2,k,\ua} 
 \;\;.\;\;
\ee

Using the commutativity of $C_{i,k,\de}^{\dagger}$'s and generalizing
the previous equations, we obtain 
($\langle B \rangle =\langle \Psi | B | \Psi \rangle)$
\bea
n_{i,k,\si}  = \langle \cd_{i,k,\si} \; c_{i,k,\si} \rangle
= |\g_{i k \si,i}|^2 \;
( |v_{i,k}|^2 + |w_{i,k,\si}|^2+|s_{i,k}|^2 +|t_{i,k,\si}|^2)
+|\g_{i k \si,j}|^2 \; (|s_{j,k}|^2 + |t_{j,k,\si}|^2)
\;\;, \\
\ze_{k,\si} = \langle \cd_{i,k,\si}  \; c_{j,k,\si} \rangle
= - \g_{i k \si,i}^* \; \g_{j k \si,i} \;
(w_{i,k,\si}^* \; t_{i,k,\si} +v_{i,k}^* s_{i,k}) 
-\g_{i k \si,j}^* \; \g_{j k \si,j} \;
(w_{j,k,\si} \; t_{j,k,\si}^* + v_{j,k} s_{j,k}^* )  \; \; ,  \\
\la_{k,\si} = \langle c_{2,-k,\si} c \; _{1,k,\si} \rangle
= \g_{1 k \si,1} \; \g_{2 -k \si,1} \; u_{1,k}^* \;  t_{1,k,\si}
- \g_{1 k \si,2} \; \g_{2 -k \si,2} \; u_{2,k}^* \;  t_{2,k,\si} \; \; ,\\
g_{k,\si} = \langle c_{2,-k,-\si}  \; c_{1,k,\si} \rangle
=  \g_{1 k \si,1} \; \g_{2 -k -\si,1} \;u_{1,k}^* \;s_{1,k} 
-\g_{1 k \si,2} \; \g_{2 -k -\si,2} \; u_{2,k}^* \;s_{2,k}   \;\; , \\
b_{i,k,\si} = \langle c_{i,-k,-\si}  \; c_{i,k,\si} \rangle
=  \g_{i k \si,i} \; \g_{i -k -\si,i} \; u_{i,k}^* \; v_{i,k} \; \; , \;\;
 d_{i,k,\si} = \langle c_{i,-k,\si}  \; c_{i,k,\si} \rangle
= \g_{i k \si,i} \; \g_{i -k \si,i}  \; u_{i,k}^* \; w_{i,k,\si} 
 \;\; , \;\;
\eea
with $(i,j)=(1,2),(2,1)$.

A general Hamiltonian for two fermion species interacting via intra-species 
potentials $V_{1,2}$ and via an inter-species potential $F_q$, 
and hybridizing via $h_k$, is
\bea
H = \sum_{i,k,\si} \xi_{i,k,\si} \;  \; \cd_{i,k,\si} c_{i,k,\si}
+ \sum_{k,\si} h_k  
\left( \cd_{1,k,\si} c_{2,k,\si} + \cd_{2,k,\si} c_{1,k,\si} \right)
\\
+ \frac{1}{2} \sum_{i,k,p,q,\si,\si'} V_{i,q} \; 
\cd_{i,k+q,\si} \cd_{i,p-q,\si'} c_{i,p,\si'}c_{i,k,\si}
+ \sum_{k,p,q,\si,\si'} F_q \; \cd_{1,k+q,\si} \cd_{2,p-q,\si'} 
c_{2,p,\si'}c_{1,k,\si}
\;\;,  \nonumber 
\eea
with $i=1,2$, $\xi_{i,k,\si}=\ep_{i,k,\si}-\mu_{i,\si}$ and 
$\mu_{i,\si}$ the chemical potential.

Considering $\Psi$ above, we have 
for $\langle H\rangle =\langle \Psi|H|\Psi\rangle $,
\bea
\langle H\rangle  = \sum_{i,k,\si} \xi_{i,k,\si} \; n_{i,k,\si} 
+ \sum_{k,\si}  \; h_k \; \big( \ze_{k,\si} +  \ze_{k,\si}^* \big) 
-\frac{1}{2} \sum_{i,k,p,\si} V_{i,k-p}\; n_{i,k,\si} \;n_{i,p,\si}
+ \frac{1}{2} \sum_{i,k,p,\si} V_{i,q=0}\; n_{i,k,\si} \;n_{i,p,\si}
\\
+ \frac{1}{2} \sum_{i,k,p,\si} V_{i,k-p}\; \big( b_{i,k,\si} b_{i,p,\si}^* \; 
+ d_{i,k,\si}^* d_{i,p,\si}   \big)
-\sum_{k,p,\si} F_{k-p} \; \ze_{k,\si} \; \ze_{p,\si}^* 
+  F_{q=0} \sum_{k,p,\si} \; n_{1,k,\si} \;n_{2,p,\si}   \nonumber \\
+ \sum_{k,p,\si} F_{k-p} \; \big( \la_{k,\si} \; \la_{p,\si}^*
+ g_{k,\si} \; g_{p,\si}^* \big)
\nonumber \;\;,  
\eea
with $(i,j)=\{(1,2),(2,1)\}$. The first term in the second line is exactly
the usual BCS-like pairing term, and the last term is 
the equivalent inter-species pairing term due to $F_q$.
Allowing for pairs with non-zero total momentum in $|\Psi\rangle $,
as shown in \cite{gk}, yields additional terms in $\langle H\rangle $.

The minimization procedure for $\langle H\rangle$ and the finite temperature
extension proceed as shown in \cite{gk}.
In general, expanding the Hilbert space of $|\Psi\rangle $ via the 
inclusion of more pairing correlations than the ones shown, may
lead to a further reduction of the ground state energy.

\vspace{0.3cm}

\vspace{.3cm}

$^*$ E-mail address : kast@iesl.forth.gr , giwkast@gmail.com

\end{document}